# Structural, electronic properties and Fermi surface of ThCr$_2$Si$_2$-type tetragonal KFe$_2$S$_2$, KFe$_2$Se$_2$, and KFe$_2$Te$_2$ phases as parent systems of new ternary iron-chalcogenide superconductors

I.R. Shein • A.L. Ivanovskii

**Abstract** The first principles FLAPW-GGA method was used for comparative study of the structural and electronic properties of three related tetragonal ThCr$_2$Si$_2$-type phases KFe$_2$Ch$_2$, where Ch are S, Se, and Te. The main trends in electronic bands, densities of states, and Fermi surfaces for $A$Fe$_2$Ch$_2$ were analyzed in relation to their structural parameters. We found that anion replacements (S↔Se↔Te) produce no critical changes in the electronic structure of KFe$_2$Ch$_2$ phases. On the other hand, our analysis of structural and electronic parameters for hypothetical KFe$_2$Te$_2$ allows us to propose this system as a perspective parent phase for search of new iron-chalcogenide superconducting materials.

**Keywords:** Ternary iron chalcogenides KFe$_2$S$_2$, KFe$_2$Se$_2$, KFe$_2$Te$_2$, Structural, electronic properties, FLAPW-GGA

## 1 Introduction

The discovery [1] of superconductivity in close proximity to magnetism in very intriguing iron-based layered materials (with maximal $T_C$ to 55K) aroused tremendous interest and motivated active research that resulted in finding of several related groups of Fe-based superconductors (SCs) such as $A$Fe$Pn$ (111-type, $Pn$ = pnictogens), $A$Fe$_2$Pn$_2$ ($A$ = alkali or alkali-earth metals, 122-type), $Ln$Fe$Pn$O(F) ($Ln$ = 4$f$ metals, 1111-type), and Sr$_4$$M_2$Fe$_2$$Pn_2$O$_6$ ($M$ = 3$d$ metals, 21113-type). All these **iron-pnictide** phases include 2D-like [Fe$_2$$Pn_2$] blocks, which are separated by $A$ atomic sheets or by more complex oxide blocks, which act as so-called charge reservoir layers, reviews [2-7]. Besides, until recently, a unique group of binary **iron-chalcogenide** Fe$Ch$ ($Ch$ = chalcogens) superconducting materials was known (11-type, without charge reservoir layers) - such as FeSe, FeTe, FeSe$_x$Te$_y$, review [8].

Very recently, the newest family of 122-type ternary **iron-chalcogenide** SCs $A$Fe$_2$Se$_2$ ($A$ = alkali metals or Tl; with $T_C$ to 35K) has been discovered, see [9,10]. These materials adopt a tetragonal ThCr$_2$Si$_2$-type structure, in which Fe atoms form a square lattice, whereas Se atoms are located at the apical sites of tetrahedrons {FeSe$_4$}; in turn, these tetrahedrons form quasi-two-dimensional blocks [Fe$_2$Se$_2$]. In general, the structure of $A$Fe$_2$Se$_2$ can be schematically described as a stacking of $A$ sheets and [Fe$_2$Se$_2$] blocks in the sequence: ...[Fe$_2$Se$_2$]/$A$/[Fe$_2$Se$_2$]/$A$/[Fe$_2$Se$_2$]... as shown in Fig. 1.

All these materials show high chemical flexibility to a large variety of constituent elements and chemical substitutions, together with high structural flexibility. Therefore,

I.R. Shein • A.L. Ivanovskii (✉)
Institute of Solid State Chemistry, Ural Branch of the Russian Academy of Sciences, 620990, Ekaterinburg, Russia
E-mail: ivanovskii@ihim.uran.ru.

numerous attempts to establish correlations between structural parameters *versus* various properties of these materials were undertaken.

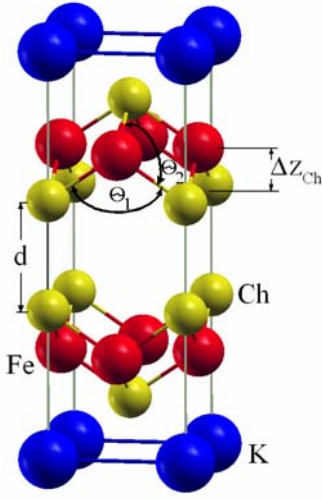

**Fig. 1** (*Color online*). Crystal structure of $KFe_2Ch_2$ phases. Some structural parameters (*Ch-Ch* bond length, *Ch*-Fe-*Ch* bond angles, and anion height, *see text*) are show.

So, for iron-pnictides the correlations between the electronic properties and $T_C$ *versus*:
(1) the *a* axis length, *i.e.* the distance between Fe atoms in the square lattice;
(2) the As-Fe-As bond angles, *i.e.* the deviation from the angles for regular {FeAs$_4$} tetrahedrons;
(3) the orthorhombic distortion $\delta = (c-a)/(c+a)$ of tetragonal lattice;
(4) the so-called anion height $\Delta z_a$, *i.e.* the distance of *Pn* atoms from the Fe plane; and
(5) the thickness of the charge reservoir layers, *i.e.* *c*-axis lattice parameters, have been extensively discussed, reviews [2-7].

Some of these correlations (between $T_C$ and anion height $\Delta z_a$ and As-Fe-As bond angles) were used recently [11] for discussion of the properties of newest $AFe_2Se_2$ SCs.

Besides, exclusively for $AFe_2Se_2$ phases, an additional possible correlation between superconductivity and *c*-axis lattice parameters was proposed [12]. The point is that for these species (which are usually cation-deficient materials) two *c*-axis lattice parameters $c_1$ and $c_2$ were observed within XRD experiments [12]. This is related to inhomogeneous distribution of alkali ions inside *A* sheets, and superconductivity exists within a limited range of $c_{1,2}$ parameters.

Generally, for $AFe_2Se_2$ phases, the tuning of the structural parameters may be achieved owing to ***cation*** substitutions (in *A*- or (and) iron sites) or owing to ***anion*** substitutions in Se sites.

The available results demonstrate that the substitutions in *A* sites influence first of all the parameter *c* (*i.e.* the [Fe$_2$Se$_2$]-[Fe$_2$Se$_2$] inter-blocks distance; in the sequence $KFe_2Se_2 \rightarrow RbFe_2Se_2 \rightarrow CsFe_2Se_2$, the *c* parameters increase from ~ 14.03 Å to ~ 15.28 Å as the radii of K, Rb, Cs ions increase: $R(K)$ = 1.51 Å < $R(Rb)$ = 1.63 Å < $R(Cs)$ = 1.78 Å), whereas the parameter *a* varies only slightly.

A much more effective way for tuning the geometry of superconducting [Fe$_2$Se$_2$] blocks will be substitutions in Fe or Se sites. Indeed, very recently the first ternary 122-like iron-sulfide phase was synthesized [13,14]. The obtained lattice parameters depend on the stoichiometry of samples (*a* = 3.753 Å and *c* = 13.569 Å for $K_xFe_{2-y}S_2$ [13] and *a* = 3.745 Å and *c* = 13.578 Å for $K_xFe_{1.7}S_2$ [14]), but were found significantly smaller than those for $K_xFe_2Se_2$ (*a* = 3.914 Å and *c* = 14.037 Å [9]) owing to substitution of S ($R^{atom}$ = 1.22 Å) for Se ($R^{atom}$ = 1.60 Å).

Here, a question arises whether the replacements in anion sites can cause critical changes on the electronic properties of ternary ***iron-chalcogenide*** SCs and how the structural parameters of these materials correlate with their band structure and Fermi surface (FS) topology.

In this work, we will try to clarify in details the changes in the band structure and FS topology *versus* possible replacements in anion sites for ternary iron-chalcogenide SCs using a series of isostructural and isoelectronic phases (synthesized $KFe_2S_2$, $KFe_2Se_2$ and hypothetical $KFe_2Te_2$) as an example. Let us also note that as distinct from the $KFe_2Se_2$ phase, for which the electronic structure was examined [15-17], there are no reports on the electronic band structure and Fermi surface topology for $KFe_2S_2$ and

KFe$_2$Te$_2$.

## 2 Models and computational aspects

We have examined a series of ternary iron-chalcogenide phases with the nominal compositions KFe$_2$Ch$_2$ (Ch = S, Se, and Te) and a tetragonal ThCr$_2$Si$_2$-type structure (space group *I*4/*mmm*; #139), Fig. 1. The atomic positions are K: 2*a* (0, 0, 0), Fe: 4*d* (0, ½, ¼), and Ch: 4*e* (0, 0, $z_{Ch}$), where $z_{Ch}$ is the so-called internal coordinate.

Note again that the sulfur- and Se-containing KFe$_2$Ch$_2$ phases have been successfully prepared [9-14], whereas KFe$_2$Te$_2$, as far as we know, still remains a hypothetical system. On the other hand, among the binary FeCh species, which are usually used in well documented [9-14] two-step routes for synthesis of KFe$_2$Ch$_2$ (Ch = S, Se) phases as precursors, FeTe and FeSe$_x$Te$_y$ have been successfully prepared [8]. This allows one to expect synthesis of Te-containing *A*Fe$_2$Ch$_2$ phases in the near future.

Since we will focus on the electronic properties of KFe$_2$Ch$_2$ as depending on the structural factors, let us note also that the experimentally measured lattice parameters (*a* and *c*) are larger than those obtained within *ab initio* calculations [15-17]. The reason is that *A*Fe$_2$Ch$_2$ phases are usually cation-deficient species such as charge-balanced K$^{1+}_{0.8}$[Fe$^{2+}_{1.6}$Se$^{2-}_{2}$]$^{0.8-}$.

Therefore, we have performed two sets of calculations. Firstly, we made full structural optimization of the ideal KFe$_2$Ch$_2$ phases both over the lattice parameters and the atomic positions including the internal coordinate $z_{Se}$, and then the electronic properties were calculated. Secondly, similar calculations were continued using the available experimental structural parameters for the compositions K$_x$Fe$_{2-y}$Se$_2$ [9] and K$_x$Fe$_{2-y}$S$_2$ [13].

All our calculations were carried out by means of the full-potential method within mixed basis APW+lo (LAPW) implemented in the WIEN2k suite of programs [18]. The generalized gradient correction (GGA) to exchange-correlation potential in the PBE form [19] was used. The plane-wave expansion was taken to $R_{MT} \times K_{MAX}$ equal to 8, and the *k* sampling with 12×12×12 *k*-points in the Brillouin zone was used. The hybridization effects were analyzed using the densities of states (DOSs), which were obtained by a modified tetrahedron method [20].

## 3. Results and discussion.

### 3.1. Structural data.

The calculated equilibrium lattice parameters (as well as some other structural data: internal coordinates $z_{Ch}$, Fe-Fe, Fe-Ch, and Ch-Ch distances *d*, Ch-Fe-Ch bond angles $\Theta_{1,2}$, and anion height $\Delta z$, see Fig. 1) for the examined KFe$_2$Ch$_2$ phases are presented in Table 1.

**Table 1**
Structural parameters for KFe$_2$Ch$_2$ (where Ch = S, Se, and Te): lattice constants (*a*, *c*, Å), internal coordinate ($z_{Ch}$), bond length (*d*, Å), bond angles ($\Theta$,°), and anion height ($\Delta z$, Å) as obtained from FLAPW-GGA.

| phase | *a* | *c* | $z_{Ch}$ |
|---|---|---|---|
| KFe$_2$S$_2$ | 3.7207 | 13.1697 | 0.3404 |
| KFe$_2$Se$_2$ | 3.8608 | 13.8369 | 0.3452 |
| KFe$_2$Te$_2$ | 4.0694 | 14.7534 | 0.3475 |
| phase | *d*(Fe-Ch) | *d*(Fe-Fe) | *d*(Ch-Ch) |
| KFe$_2$S$_2$ | 2.2110 | 2.6310 | 4.2030 |
| KFe$_2$Se$_2$ | 2.3370 | 2.7299 | 4.3196 |
| KFe$_2$Te$_2$ | 2.4881 | 2.8704 | 4.4906 |
| phase | $\Theta_1$ | $\Theta_2$ | $\Delta z$ |
| KFe$_2$S$_2$ | 114.58 | 106.89 | 1.4276 |
| KFe$_2$Se$_2$ | 111.41 | 108.52 | 1.3165 |
| KFe$_2$Te$_2$ | 109.54 | 109.33 | 1.4461 |

The calculated parameters allow us to make the following conclusions.

In the sequence KFe$_2$S$_2$ → KFe$_2$Se$_2$ → KFe$_2$Te$_2$, both the *a* and *c* parameters increase; this result can be easily explained by considering the atomic radii of Ch atoms: $R^{atom}$(S) = 1.22 Å < $R^{atom}$(Se) = 1.60 Å < $R^{atom}$(Te) = 1.70 Å.

At the same time, when going from $KFe_2S_2$ to $KFe_2Te_2$, the parameter $c$ grows by about 10.7%, whereas the parameter $a$ grows by about 9.4 %, and the $c/a$ ratio increases from 3.54 to 3.63. Thus, the *anisotropic deformation* of the crystal structure occurs, which is caused by strong *anisotropy of interatomic bonds*, *i.e.* by strong bonds inside [$Fe_2Ch_2$] blocks *versus* relatively weak coupling between adjacent [$Fe_2Ch_2$]/[$Fe_2Ch_2$] blocks [21].

Next, inside each [$Fe_2Ch_2$] block, the anion replacements also result in *anisotropic* changes in Fe-Fe and Fe-$Ch$ inter-atomic distances. So, when going from $KFe_2S_2$ to $KFe_2Te_2$, the $d$(Fe-$Ch$) distances grow by about 12.5%, whereas the $d$(Fe-Fe) distances grow only by about 9.1 %. Thus, we can conclude that the anion replacements result in anisotropic deformation both of [$Fe_2Ch_2$] blocks and a crystal as a whole.

In this context, of interest are the changes in the aforementioned structural parameters (namely, the $Ch$-Fe-$Ch$ bond angles and the anion height $\Delta z_{Ch}$), which are often used for correlations with electronic and superconducting properties of iron-based materials. For $KFe_2S_2$, the S-Fe-Se bond angles ranging from $\Theta_1$ = 114.58º to $\Theta_2$ =106.89º are the furthest from the ideal tetrahedron angle of 109.47º, resulting probably in non-superconducting behavior of this phase [13,14]. Further, in the sequence $KFe_2S_2 \rightarrow KFe_2Se_2 \rightarrow KFe_2Te_2$, the distortions of {Fe$Ch_4$} tetrahedrons decrease owing to an increase in $\Theta_2$ and a decrease in $\Theta_1$ angles; for $KFe_2Te_2$, these angles become very close to the ideal tetrahedron angles: $\Theta_1$ = 109.54º and $\Theta_2$ =109.33º. Thus, according the the aforementioned indicator, $KFe_2Te_2$ should be assumed as a promising superconducting material.

On the other hand, the anion height $\Delta z_{Ch}$ in the sequence $KFe_2S_2 \rightarrow KFe_2Se_2 \rightarrow KFe_2Te_2$ varies non-monotonically (Table 1) becoming smaller (for $KFe_2Se_2$) or greater (for $KFe_2S_2$ and $KFe_2Te_2$) than the "optimal" factor $\Delta z$ = 1.38 Å [22].

## 3.2. Electronic properties and Fermi surface.

The energy bands, Fermi surfaces, as well as total and site-projected *l*-decomposed densities of states (DOS) for the examined $KFe_2Ch_2$ phases, which have been calculated for their equilibrium geometries, are shown in Figs. 2-4, respectively.

According to the results obtained, their electronic spectra appear very similar. Therefore, we will discuss their peculiarities in more detai, using $KFe_2S_2$ as an example.

Figure 2 shows the near-Fermi band structure of $KFe_2S_2$ along the selected high-symmetry lines within the first Brillouin zone of the tetragonal crystal. It is seen that there is no gap at $E_F$, and the Fermi level is crossed by Fe 3*d*-like bands, which indicates that the electrical conductivity of this phase should be metallic-like. The near-Fermi bands demonstrate a complicated "mixed" character: simultaneously with quasi-flat bands along Γ-Z, a series of high-dispersive bands intersects the Fermi level.

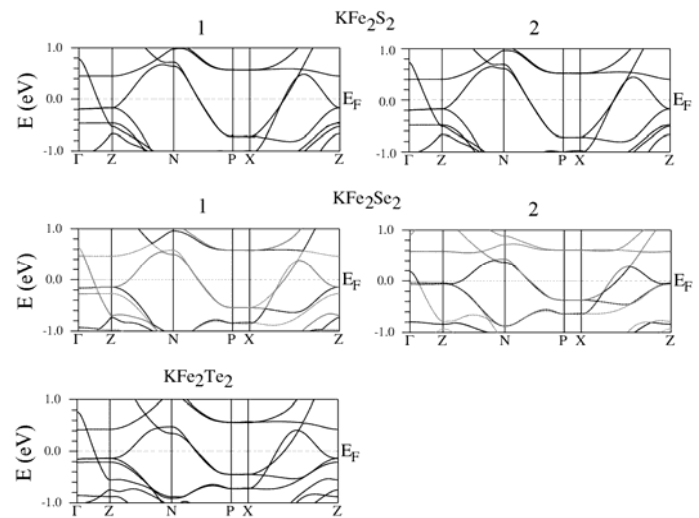

**Fig. 2** The near-Fermi bands for $KFe_2S_2$, $KFe_2Se_2$, and $KFe_2Te_2$ phases with optimized lattice constants (1) and for experimental parameters (2).

These features yield a multi-sheet FS, which consists (Fig. 3) of two quasi-two-dimensional (2D) electron-like sheets in the corners of the Brillouin zone, and the closed disconnected electron-like pockets (around *Z*) - instead of cylinder-like hole-like sheets for 122 Fe-*Pn* materials [2-7]. As a result, the Fermi surface nesting effect in $KFe_2S_2$ (as well as for $KFe_2Se_2$ and $KFe_2Te_2$, Fig. 3) is absent, implying that the inter-pocket scattering ($s_\pm$-type pairing symmetry) is absent in the superconducting state. For $KFe_2Se_2$ and $KFe_2Te_2$, their near-Fermi band pictures and Fermi surfaces are very similar to those for $KFe_2S_2$, Figs. 2,3. Only small differences (in the distances of quasi-flat bands from the Fermi level) are visible; thus, the anion replacements have no critical influence on the near-Fermi electronic bands and the FS.

negligible, *i.e.* in $KFe_2S_2$ these atoms are in the form of cations $K^+$.

In the sequence $KFe_2S_2 \to KFe_2Se_2 \to KFe_2Te_2$, the overall DOS structure is preserved, but the width of the common valence band and the widths of the aforementioned subbands A and B and the gaps decrease.

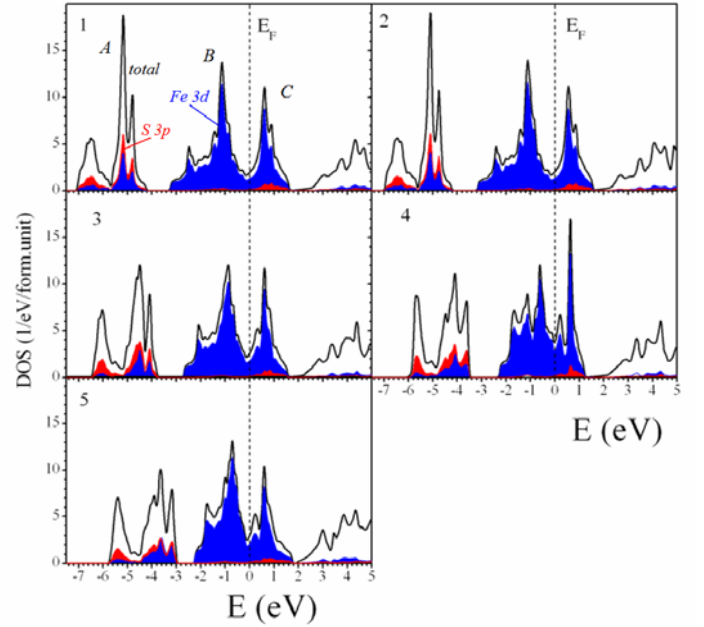

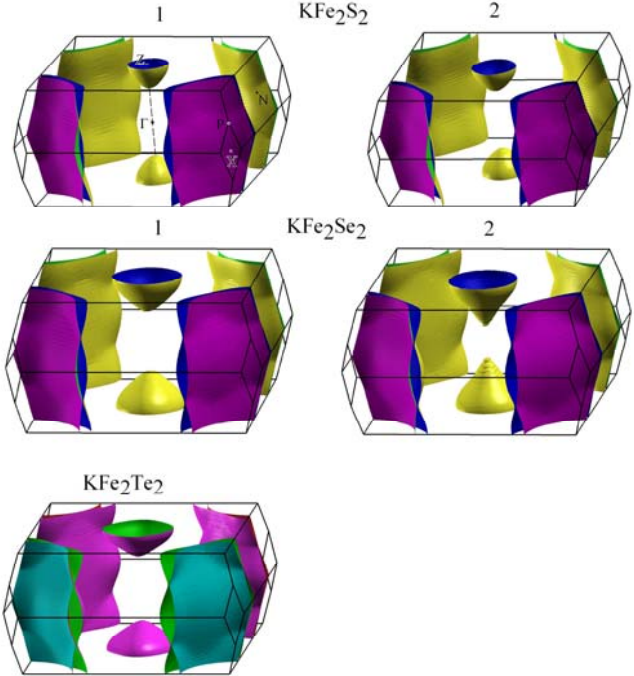

**Fig. 3** (*Color online*). The Fermi surfaces of $KFe_2S_2$, $KFe_2Se_2$, and $KFe_2Te_2$ phases with optimized lattice constants (1) and for experimental parameters (2).

**Fig. 4** (*Color online*). Total and *l*-resolved partial densities of states for $KFe_2S_2$ (1,2), $KFe_2Se_2$ (3,4), and $KFe_2Te_2$ (5) phases with optimized lattice constants (1,3,5) and for experimental parameters (2,4).

**Table 2.**
Total and partial densities of states ($N(E_F)$, states/eV·f.u.) at the Fermi level for $KFe_2Ch_2$ ($Ch$ = S, Se, Te).

| phase | | total | Fe 3$d$ | $Ch\ p$ |
|---|---|---|---|---|
| $KFe_2S_2$ | 1* | 1.939 | 1.287 | 0.114 |
| | 2 | 2.025 | 1.344 | 0.118 |
| $KFe_2Se_2$ | 1 | 2.571 | 1.772 | 0.122 |
| | 2 | 3.811 | 2.873 | 0.143 |
| $KFe_2Te_2$ | 1 | 2.810 | 1.921 | 0.102 |

\* as calculated for optimized (1) and experimental (2) lattice parameters

In the density of states of $KFe_2S_2$, the lowest bands (peak *A*, Fig. 4) lying in the range from -7.1 to -4.2 eV below the Fermi level arise mainly from mixed Fe3$d$-S3$p$ states and are separated from the near-Fermi valence bands by a gap. These bands (peak *B*) are located in the energy range from -3.1 eV to $E_F$ and are formed predominantly by Fe3$d$ states. Finally, the lowest part of the conduction band (peak *C*) is also made up basically of contributions from Fe3$d$ states with an admixture of anti-bonding S3$p$ states. Besides, the contributions from the potassium states in the occupied subbands A and B are quite

So, according to our calculations, the widths of the common valence band and the gap for $KFe_2S_2$ are about 7.1 eV and 1.1 eV *versus* 5.7 eV and 0.7 eV for $KFe_2Te_2$. Besides, obvious differences occur in the total and partial DOSs at the Fermi level, $N(E_F)$, which are shown in Table 2. Here, in the sequence $KFe_2S_2 \to$ $KFe_2Se_2 \rightarrow KFe_2Te_2$, the value of $N^{tot}(E_F)$ grows from 1.9 states/eV·f.u. for $KFe_2S_2$ to 2.8 states/eV·f.u. for $KFe_2Te_2$ owing to Fe3*d* contributions, whereas the contributions from *Ch p* states change only slightly, Table 2.

The obtained data also allow us to estimate the Sommerfeld constants for ternary $KFe_2Ch_2$ phases under the assumption of the free electron model as $\gamma = (\pi^2/3)N(E_F)k_B^2)$. The calculated $\gamma$ values increase in the sequence $KFe_2S_2$ (4.57 mJ·K$^{-2}$·mol$^{-1}$) < $KFe_2Se_2$ (6.06 mJ·K$^{-2}$·mol$^{-1}$) < $KFe_2Te_2$ (6.62 mJ·K$^{-2}$·mol$^{-1}$). Our estimations agree with available experimental data: for $KFe_{2-y}Se_2$ $\gamma^{exp}$ = 5.8 mJ·K$^{-2}$·mol$^{-1}$ [23].

Finally, in framework of the oversimplified model, which was used for (K,Cs)Fe$_2$Se$_2$ [17], rough estimations of the trends in $T_C$ were made. Here, the BCS-like expression for $T_C = 1.14\omega_D e^{-2/\lambda N(E_F)}$ was employed, and, as the authors say [17], these estimations do not necessarily imply electron-phonon pairing, as $\omega_D$ may just denote the average frequency of any other possible Boson responsible for pairing interaction (e.g. spin fluctuations). Using the calculated values of $N(E_F)$, the following tendency in $T_C$ was obtained: 3.8K ($KFe_2S_2$) < 10K ($KFe_2Se_2$) < 13.7K ($KFe_2Te_2$), which allow one to assume that the hypothetical $KFe_2Te_2$ may be a potential parent phase for the search of new ternary iron-chalcogenide SCs.

*3.3. Electronic properties versus structural parameters.*

As it was pointed out earlier, all of the synthesized $AFe_2Se_2$ phases are non-stoichiometric owing to the presence of cation vacancies both in *A* and Fe sublattices. Thus, to provide an additional insight into the influence of cation non-stoichiometry induced changes in the structural parameters on the electronic properties of the examined phases, we have calculated $KFe_2S_2$ and $KFe_2Se_2$ phases using the experimental structural parameters for $K_xFe_{2-y}Se_2$ [9] and $K_xFe_{2-y}S_2$ [13].

The results obtained are shown in Figs. 2-4 and Table 2. We see that the main peculiarities of the electronic bands and the Fermi surface topology (namely: (i) strongly anisotropic near-Fermi bands of a complicated "mixed" character: quasi-flat and high-dispersive bands intersect the Fermi level; (ii) the sheets forming the Fermi surface are exclusively of an electronic-like type; (iii) the band width narrow and $N(E_F)$ grows in the sequence $KFe_2S_2 \rightarrow KFe_2Se_2$) are the same as obtained earlier for the optimized structural parameters.

Thus, we can conclude that the observed peculiarities of the transport, magnetic, and electronic properties of the synthesized $A_xFe_{2-y}Se(S)_2$ samples [9-14] are mainly governed by non-stoichiometry effects such as the concentration of lattice vacancies, their ordering *etc.*, whereas the accompanying changes in the lattice parameters play a relatively insignificant role.

## 4 Conclusions

In summary, by means of the FLAPW-GGA approach, we have systematically studied the trends in structural and electronic properties of three related tetragonal ThCr$_2$Si$_2$-type phases $KFe_2Ch_2$, where *Ch* are S, Se, and Te.

Our main findings are as follows. Firstly, the anion replacements (S ↔ Se ↔ Te) for $KFe_2Ch_2$ phases result in anisotropic deformations both of [Fe$_2$Ch$_2$] blocks and a crystal as a whole. Secondly, the aforementioned anion replacements (as well as the variation in the lattice parameters) for *the ideal stoichiometric* $KFe_2Ch_2$ phases do not lead to any critical changes in their near-Fermi electronic bands and the FS topology. Namely, for all these systems, similar multi-sheet Fermi surfaces consist exclusively of electronic-like sheets, two of which are 2D-like sheets in the corners of the Brillouin zone, and the third is a closed pocket around *Z*.

On the other hand, our calculations reveal that in the sequence $KFe_2S_2 \rightarrow KFe_2Se_2 \rightarrow KFe_2Te_2$ the distortions of {Fe*Ch*$_4$} tetrahedrons decrease, and for $KFe_2Te_2$, the Ch-Fe-*Ch* bond angles become very close to the ideal tetrahedron angles. Besides, in this sequence, the value of the DOSs at the Fermi level grows. Both these factors are favorable for the formation of the superconducting state. Thus, these data allow us to assume that $KFe_2Te_2$ may be proposed as a potential parent phase for the search of new ternary iron-chalcogenide SCs.

**Acknowledgments**
The authors acknowledge the support from the RFBR (grants No. 09-03-00946 and No. 10-03-96008).